\documentclass[showpacs,aps,prl,twocolumn,superscriptaddress]{revtex4-1}
\usepackage{graphicx} 
\usepackage{dcolumn}
\usepackage{bm}
\usepackage{amssymb,amsmath}
\usepackage{epstopdf}
\usepackage{color}

\begin{document}
\title{Probing the semiconductor to semimetal transition in InAs/GaSb double quantum wells by magneto-infrared spectroscopy}

\author{Y. Jiang}
\affiliation{School of Physics, Georgia Institute of Technology, Atlanta, Georgia 30332, USA}
\author{S. Thapa}
\affiliation{Department of Physics, University of Florida, Gainesville, Florida 32611, USA}
\author{G. D. Sanders}
\affiliation{Department of Physics, University of Florida, Gainesville, Florida 32611, USA}
\author{C. J. Stanton}
\affiliation{Department of Physics, University of Florida, Gainesville, Florida 32611, USA}
\author{Q. Zhang}
\affiliation{Department of Electrical and Computer Engineering, Department of Physics and Astronomy, and Department of Materials Science and NanoEngineering, Rice University, Houston, Texas 77005, USA}
\author{J. Kono}
\affiliation{Department of Electrical and Computer Engineering, Department of Physics and Astronomy, and Department of Materials Science and NanoEngineering, Rice University, Houston, Texas 77005, USA}
\author{W. K. Lou}
\affiliation{SKLSM, Institute of Semiconductors, Chinese Academy of Sciences, Beijing 100083, China}
\affiliation{Synergetic Innovation Center of Quantum Information and Quantum Physics, University of Science and Technology of China, Hefei, Anhui 230026, China}
\author{K. Chang}
\affiliation{SKLSM, Institute of Semiconductors, Chinese Academy of Sciences, Beijing 100083, China}
\affiliation{Synergetic Innovation Center of Quantum Information and Quantum Physics, University of Science and Technology of China, Hefei, Anhui 230026, China}
\author{S. D. Hawkins}
\affiliation{Sandia National Laboratories, Albuquerque, New Mexico 87185, USA}
\author{J. F. Klem}
\affiliation{Sandia National Laboratories, Albuquerque, New Mexico 87185, USA}
\author{W. Pan}
\affiliation{Sandia National Laboratories, Albuquerque, New Mexico 87185, USA}
\author{D. Smirnov}
\affiliation{National High Magnetic Field Laboratory, Tallahassee, Florida 32310, USA}
\author{Z. Jiang}
\email{zhigang.jiang@physics.gatech.edu}
\affiliation{School of Physics, Georgia Institute of Technology, Atlanta, Georgia 30332, USA}

\date{\today}

\begin{abstract}
We perform a magneto-infrared spectroscopy study of the semiconductor to semimetal transition of InAs/GaSb double quantum wells from the normal to the inverted state. We show that owing to the low carrier density of our samples (approaching the intrinsic limit), the magneto-absorption spectra evolve from a single cyclotron resonance peak in the normal state to multiple absorption peaks in the inverted state with distinct magnetic field dependence. Using an eight-band Pidgeon-Brown model, we explain all the major absorption peaks observed in our experiment. We demonstrate that the semiconductor to semimetal transition can be realized by manipulating the quantum confinement, the strain, and the magnetic field. Our work paves the way for band engineering of optimal InAs/GaSb structures for realizing novel topological states as well as for device applications in the terahertz regime.
\end{abstract}

\pacs{76.40.+b, 71.70.Di, 78.20.Ls, 73.21.Fg}

\maketitle
Broken-gap InAs/GaSb double quantum wells (DQWs) have long been important in studying intriguing phenomena, ranging from excitonic resonances \cite{Kono_PRB1,Tung_PRB1}, electron-hole hybridization \cite{Nicholas_SST1,Yang_PRL,Marlow_PRL,Nicholas_PRB1}, the quantum spin Hall (QSH) effect \cite{Liu_PRL,Du_PRL2,Ihn_PRB}, helical Luttinger liquids \cite{Du_PRL3}, to the exciton insulator \cite{Du_arXiv}. The novel properties of the InAs/GaSb system stem from its peculiar band-edge alignment, in which the bottom of the InAs conduction band lies below the top of the GaSb valence band (Fig. 1(a)), resulting in the separation of electrons and holes into the two QWs. By changing the width of each QW, one can manipulate the electron and hole energy levels using the quantum confinement effect. Specifically, when the lowest electron level $E_0$ in InAs lies above the highest hole level $H_0$ in GaSb, the system is in the normal state ($E_0>H_0$). When the opposite alignment $E_0<H_0$ is achieved, the system is said to be in the inverted state. The transition from normal to inverted states occurs at the critical QW widths at which $E_0=H_0$ (Fig. 1(b)). Recently, the QSH effect has been theoretically predicted \cite{Liu_PRL} and experimentally confirmed \cite{Du_PRL2} in the inverted regime, while an exotic exciton insulator state \cite{Du_arXiv} and a giant supercurrent state \cite{Shi_JAP} have been found in the vicinity of the critical state. Therefore, precise control of band alignment of InAs/GaSb DQWs from the normal to the inverted state is crucial for future fundamental studies. In particular, a thorough understanding of strain effects may lead to a robust QSH insulator state suited for realizing Majorana fermions \cite{Du_new}.

\begin{figure}[t]
\includegraphics[width=8.5cm]{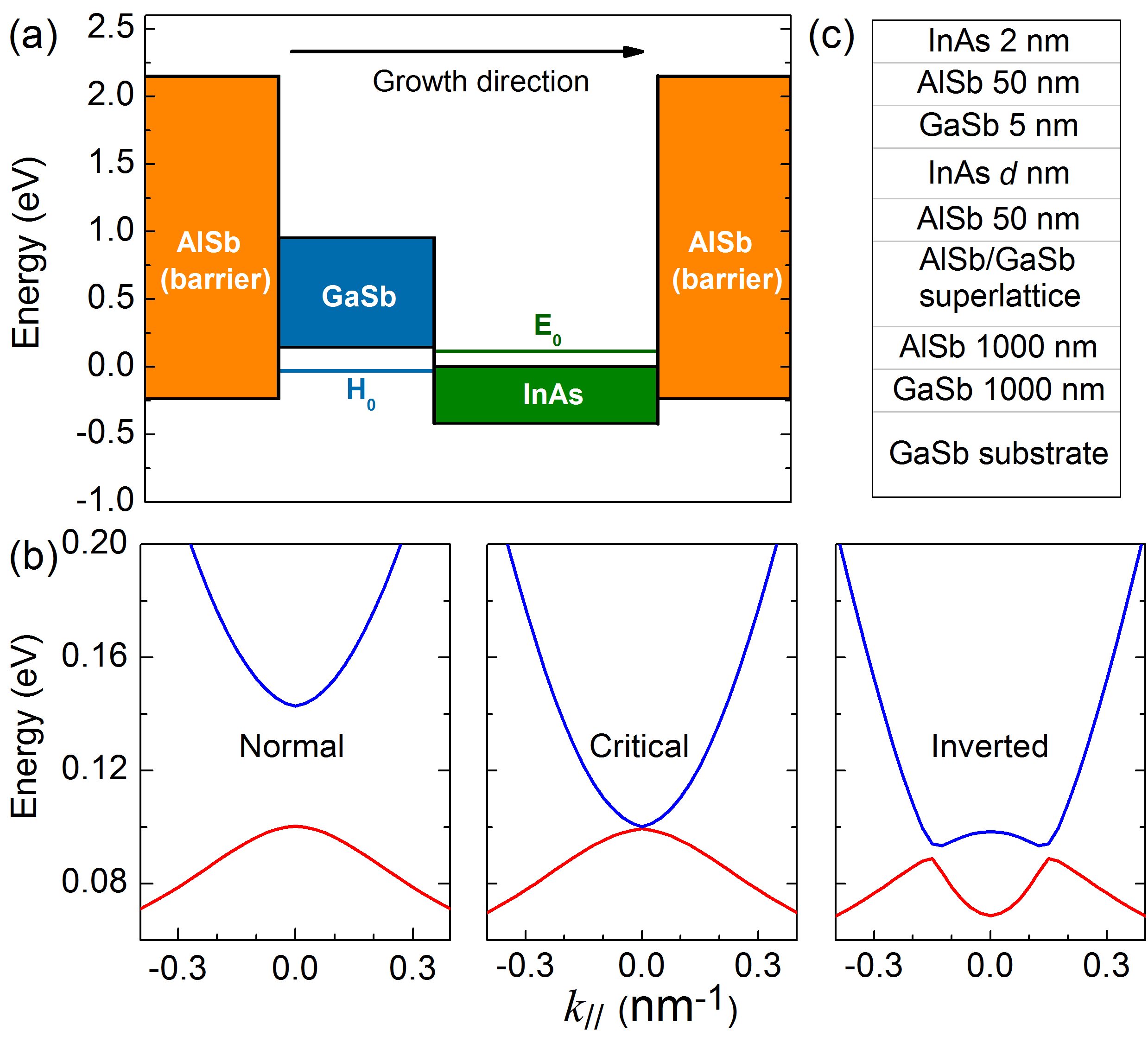}
\caption{(color online) (a) Schematic band diagram of AlSb/InAs/GaSb/AlSb QW structure. The energy zero is referenced to the conduction band edge of bulk InAs. The top and bottom of each color coded blocks indicate the energies of  the conduction and valence band edges for each material. $E_0$ and $H_0$ illustrate the lowest electron subband in InAs and the highest hole subband in GaSb, respectively, due to quantum confinement. (b) Evolution of the band alignment in InAs/GaSb DQWs, as the width $d$ of InAs QW increases (from left to right) while fixing the GaSb QW width. Blue and red curves represent the lowest electron and highest hole subbands, respectively. (c) Epi-structure of MBE grown InAs/GaSb DQW samples.}
\end{figure}

From technological perspective, InAs/GaSb-based materials are promising candidates for third-generation infrared (IR) detectors \cite{Martyniuk}, high-power light emitting diodes \cite{Prineas}, and tunnel field-effect transistors \cite{Hudait}, owing to their band engineering flexibility and the resulting low Auger recombination rates \cite{Ehrenreich}. Practical applications of the material require a complete understanding of the electronic band structure, with respect to external parameters such as strain and doping. Previous combined experimental and theoretical studies \cite{Nicholas_SST1,Yang_PRL,Marlow_PRL,Nicholas_PRB1} largely focused on the heavily inverted regime using reduced models. Quantitative investigations of the transition from the normal to the inverted state have not yet been performed.

In this Letter, we study the evolution of the electronic band structure of InAs/GaSb DQWs across the semiconductor to semimetal transition using magneto-IR spectroscopy. The observed magneto-optical modes can be explained using an eight-band $\mathbf{k\cdot p}$ model, and semi-quantitive agreement is achieved. We show that in addition to the commonly used electrostatic gate, the normal to the inverted state transition can be manipulated in a much larger parameter space via tuning the relative thickness of the QWs, the strain, and the magnetic field.

The InAs/GaSb DQW samples studied in this work were grown by molecular beam epitaxy (MBE) on GaSb (001) substrates. A schematic of the epi-structure is shown in Fig. 1(c), where the InAs/GaSb DQW structure is sandwiched between two AlSb barrier layers. To study the normal to inverted transition, we fabricated a series of five InAs/GaSb DQW samples. We fixed the width of the GaSb QW at 5 nm, while varying the InAs QW width from $d=8$ to $d=10$, 11, 13, and 15 nm. Based on our self-consistent eight-band $\mathbf{k\cdot p}$ calculation (to be elaborated later), the $d=8$ nm sample is in the normal state, the $d=10$ nm sample is close to the critical state and the $d=11$, 13, and 15 nm samples are in the inverted state. Magneto-transport measurements determined the carrier densities to be as low as $n \sim 1 \times 10^{11}$ cm$^{-2}$ \cite{Shi_JAP}, several times lower than that reported in previous studies \cite{Kono_PRB1,Tung_PRB1,Nicholas_SST1,Marlow_PRL,Nicholas_PRB1}. Therefore, our samples are close to the intrinsic limit, particularly suited for magneto-optical spectroscopy studies.

Magneto-IR spectroscopy measurements were performed in the Faraday configuration at liquid helium temperature (4.2 K) using a Bruker 80v Fourier-transform IR spectrometer. The (unpolarized) IR radiation from a mercury lamp was delivered to the sample located at the center of a 17.5 T superconducting magnet via evacuated light-pipes, and the intensity of the transmitted light was detected by a composite Si bolometer mounted beneath the sample. Normalized magneto-absorption spectra were then obtained by taking the ratio of $-T(B)/T(B=0)$, where $T(B)$ is the transmission spectrum measured at a constant magnetic field $B$. In this scenario, the intra-band (cyclotron resonance, CR) and inter-band Landau level (LL) transitions are expected to manifest themselves as a series of absorption peaks.

\begin{figure}[t]
\includegraphics[width=8.5cm]{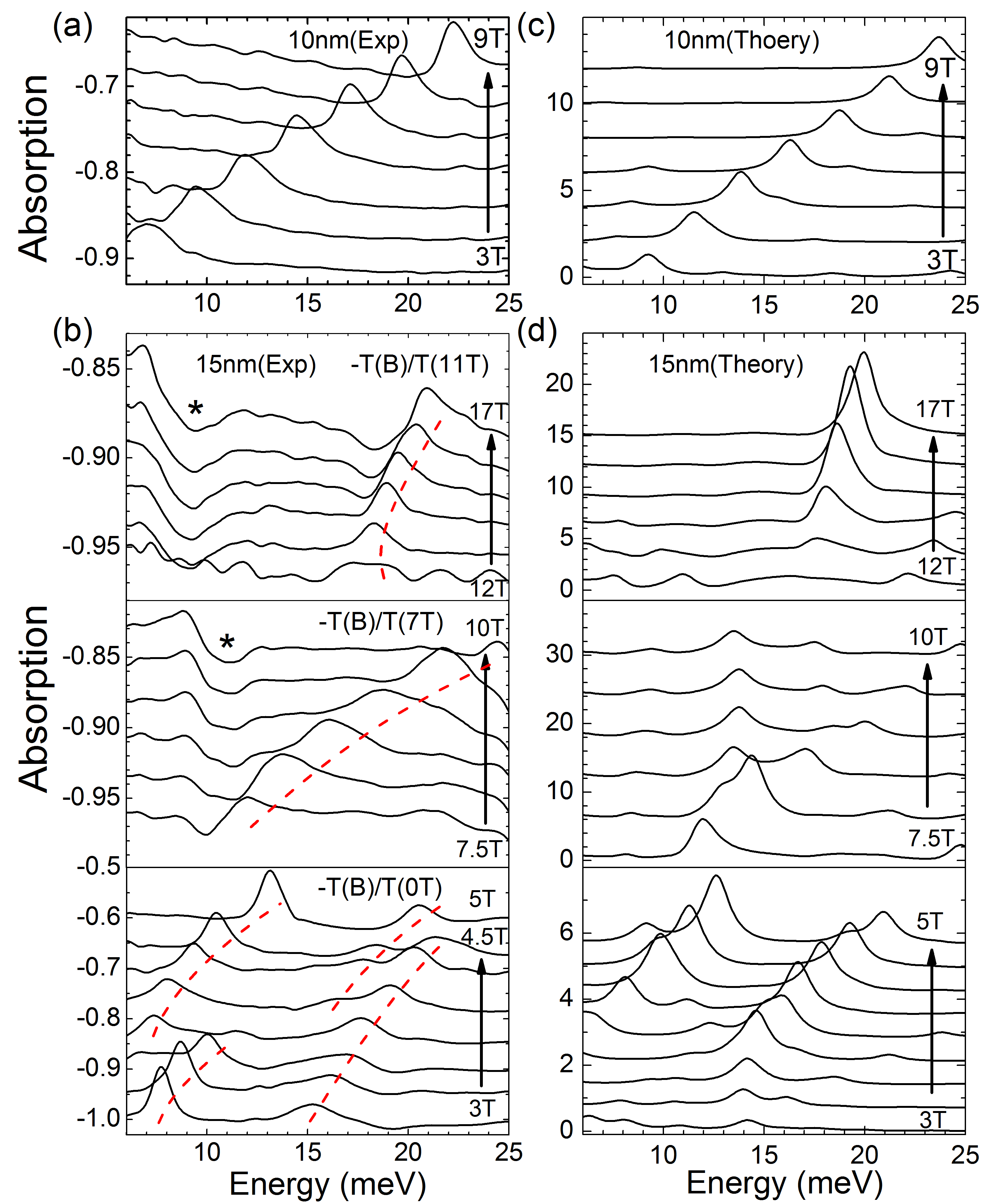}
\caption{(color online) (a) Normalized magneto-absorption spectra, $-T(B)/T(B=0)$, for the slightly inverted InAs/GaSb DQW sample ($d=10$ nm) at selected magnetic fields. (b) Normalized magneto-absorption spectra for the heavily inverted sample ($d=15$ nm) in the high-field (top panel), intermediate-field (middle panel), and low-field (bottom panel) regions. For best presentation, the spectra are normalized to 11 T, 7 T, and 0 T, respectively, for each panel. The star symbols point to $B$-independent spectral features originating from the normalization process. The dashed lines indicate the major absorption peaks observed in the experiment. (c,d) Calculated magneto-absorption spectra using the eight-band PB model, in comparison with the experimental results in (a) and (b). In all panels, the spectra are offset vertically for clarity.}
\end{figure}

Figure 2(a) shows normalized magneto-absorption spectra, $-T(B)/T(B=0)$, for the $d=10$ nm DQW sample at selected magnetic fields. Here, the spectra exhibit a single symmetric line, indicative of a single absorption peak within the energy range of our measurements. The spectral lineshape becomes asymmetric and substantially broadened in the $d=11$ nm sample, as shown in Fig. S1 in the Supplemental Material \cite{SM}. As the InAs QW width increases further to $d=13$ nm (Fig. S2) and $d=15$ nm (Fig. 2(b)), the InAs/GaSb DQWs enters a heavily inverted regime and multiple absorption peaks appear. Experimentally, one can identify four peaks for the case of $d=15$ nm in Fig. 2(b) in the low-field region, $B\le5$ T. The two lower energy peaks are CR-like, $E_{CR} \propto B$, while the higher energy ones can be attributed to inter-band LL transitions. Interestingly, the CR-like peaks deviate from their linear-in-$B$ dependence at higher magnetic field (better seen in Fig. 4(d)), suggestive of LL crossing/anti-crossing. This behavior is consistent with that theoretically predicted in Ref. \cite{Chao_PRB2}. In the middle and upper panels of Fig. 2(b), one can find another two absorption peaks with distinct $B$-dependence. In particular, the high-field peak exhibits very weak $B$-dependence, similar to that reported for InAs/GaSb superlattices in Ref. \cite{Nicholas_SST1}. The observation of a total of six absorption peaks in the $d=15$ nm DQW sample (Fig. 4(d)) can be attributed to its low carrier density (approaching the intrinsic limit). However, the literature lacks a quantitative model that can interpret all these peaks.

To explain how band inversion changes the LL transitions in InAs/GaSb DQWs, we employ an eight-band Pidgeon-Brown (PB) model \cite{Pidgeon_PR} that includes the interaction between the conduction and valence bands, the Zeeman energy, and the effect of spin-orbit coupling. The PB model is based on an eight-band $\mathbf{k\cdot p}$ method, described in the literature \cite{Chao_PRB1,Chao_JPCM1,Chao_PRB2,Kai_PRB1,Kai_APL,Zhou_PRB}, which has been successfully used for achieving semi-quantitative understanding of the electronic and magneto-optical properties of narrow-gap semiconductors \cite{Stanton_PRB1,Stanton_PRB2,Stanton_JAP,Buhmann_PRB1}. We paid special attention to two important effects: (1) the effect of strain due to the lattice mismatch among different QW layers, and (2) the effect of charge transfer through the InAs/GaSb interface. We assume that the system is under pseudomorphic strain and all the QW in-plane lattice constants are pinned to the in-plane lattice constant of the GaSb substrate. The strain gives rise to an energy shift of a few tens of meV and drives the system towards a more inverted band alignment. This effect is significant for all the DQW samples studied. The charge transfer effect, on the other hand, is appreciable only in the inverted state \cite{Esaki} when the valence band of GaSb overlaps with the conduction band of InAs. This overlap leads to charge redistribution across the InAs/GaSb interface, which consequently modifies the potential profile in DQWs. The charge transfer effect counters, or even overpowers, the strain effect in heavily inverted InAs/GaSb samples. It can be accommodated by solving the $\mathbf{k\cdot p}$ equations iteratively until the potential profile reaches convergence. Further detailed information (including an improved self-consistent algorithm) can be found in the Supplemental Material \cite{SM}.

In the presence of a magnetic field, the total effective mass Hamiltonian of InAs/GaSb DQWs can be written as 
\begin{equation}
H=H_L+H_Z+H_S+H_C,
\label{Ham}
\end{equation}
where $H_L$, $H_Z$, $H_S$, and $H_C$ are the Landau, Zeeman, strain, and confinement Hamiltonian, respectively. Following the PB formalism, one can then solve the Schr\"{o}dinger equation with a convenient set of envelope functions $F_{p,\nu}$ in the axial approximation,
\begin{equation}
H_p F_{p,\nu} = E_{p,\nu} F_{p,\nu},
\label{Sch}
\end{equation}
where the integer $p$ is the PB manifold index and $p\geq-1$, and the integer $\nu$ labels the eigenvectors/eigenenergies belonging to the same index $p$. We note that the Landau Hamiltonian $H_L$ in Eq. (\ref{Ham}) is $p$-dependent, whereas $H_Z$, $H_S$, and $H_C$ are independent of $p$. The explicit expressions of these terms are given in Refs. \cite{Stanton_PRB1,Stanton_PRB2}. Next, the matrix eigenvalue equations of Eq. (\ref{Sch}) can be solved separately for each allowed value of $p$ and the eigenenergy $E_{p,\nu}$ is obtained as a function of magnetic field.

The calculated energy levels and Fermi energy $E_F$ are plotted as a function of magnetic field in Fig. 3 and Fig. S3 for all the DQW samples studied. The PB manifold is color coded based on the index $p$, and for simplicity we only show the lowest levels, \textit{i.e.}, when $p=-1$, 0, 1, and 2. As one can clearly see in Fig. 3, the band structure of InAs/GaSb DQWs exhibits a transition from the normal to the inverted state with increasing InAs QW width $d$. In the normal state (Fig. 3(a)), the electron (hole) levels reside in the conduction (valence) band and the energy of each level shows a monotonic $B$-dependence without any crossing. In contrast, when the band is inverted (Figs. 3(b-d)), hole-like levels may exist at the bottom of the conduction band while electron-like levels appear at the top of the valence band. These levels inevitably cross or anti-cross each other at sufficiently high magnetic field, leading to multiple magneto-absorption peaks. The magnitude of the crossing/anti-crossing magnetic field therefore characterizes the degree of band inversion. One can see from Fig. 3 that the $d=10$ nm DQW sample is only slightly inverted, whereas the $d=13$ nm and 15 nm samples are heavily inverted. In addition, we note that even for a heavily inverted band structure, the electron-like levels in the valence band can be lifted above all the hole-like levels, at sufficiently high magnetic fields, driving the system to the normal state \cite{Kono_PRB1,Comment2_theory,Comment2_Nicholas,Comment2_proceeding}. We will return to this magnetic field driven semimetal to semiconductor transition in the context of Fig. 5.

\begin{figure}[t]
\includegraphics[width=8.5cm] {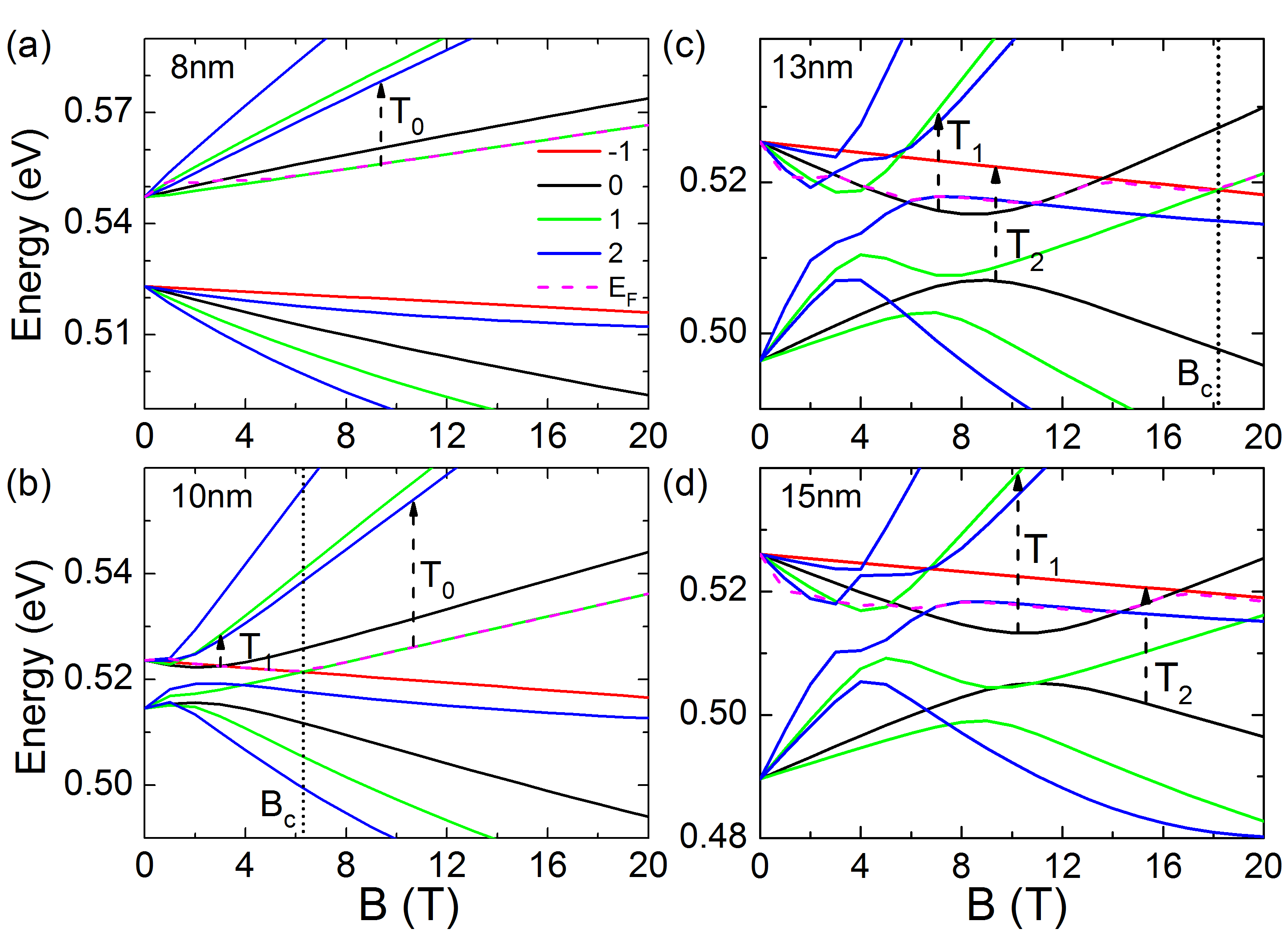}
\caption{(color online) Calculated energy levels as a function of magnetic field for (a) $d=8$ nm, (b) 10 nm, (c) 13 nm, and (d) 15 nm InAs/GaSb DQW samples. The PB manifold is color coded based on the index $p$, and the dashed line shows the evolution of $E_F$ as a function of magnetic field. The dashed arrows indicate the major transitions, $T_0$, $T_1$, and $T_2$, commonly observed in our samples. The dotted lines mark the onset ($B_c$) of the magnetic field driven transition from the inverted to the normal state.}
\end{figure}

In order to calculate the magneto-optical absorption spectra, we first determine $E_F$ using a standard routine described in Ref. \cite{Stanton_PRB1} assuming an electron density of $n=0.9 \times 10^{11}$ cm$^{-2}$. Second, we compute the magneto-absorption coefficient using the wavefunctions obtained from the PB model and Fermi's golden rule \cite{Stanton_PRB1,Stanton_PRB2}. The full width at half maximum (FWHM) is taken to be 0.8 meV in our calculation, estimated from the experimental data. The deduced selection rule reads $\Delta p=\pm1$, where $+$ ($-$) denotes electron (hole) like transitions. Figures 2(c,d) show the calculated magneto-absorption spectra for the slightly inverted ($d=10$ nm) and heavily inverted ($d=15$ nm) DQW samples, in comparison with the experimental data in Figs. 2(a,b). The same normalization method is applied to both the experimental and theoretical results. Here, as one can see, the calculated spectra capture all the absorption peaks observed in the experiment, although the relative strength between these peaks does not exhibit a perfect match. Better understanding of the data can be attained using the manifold-resolved magneto-absorption spectra, some examples of which are shown in Fig. S4. Manifold-resolved calculations help assign a specific transition to each absorption peak observed in our experiment. For example, the CR peak observed in the $d=8$ sample can be attributed to the $T_0$ transition from a $p=1$ level to a $p=2$ level ($1\rightarrow2$), as illustrated in Fig. 3(a). The manifold-resolved results are summarized in Fig. 4, where intra-band (inter-band) transitions are color coded in red (blue). Semi-quantitative agreement between the theory and experiment is reached \cite{note1}.

\begin{figure}[b]
\includegraphics[width=8.5cm]{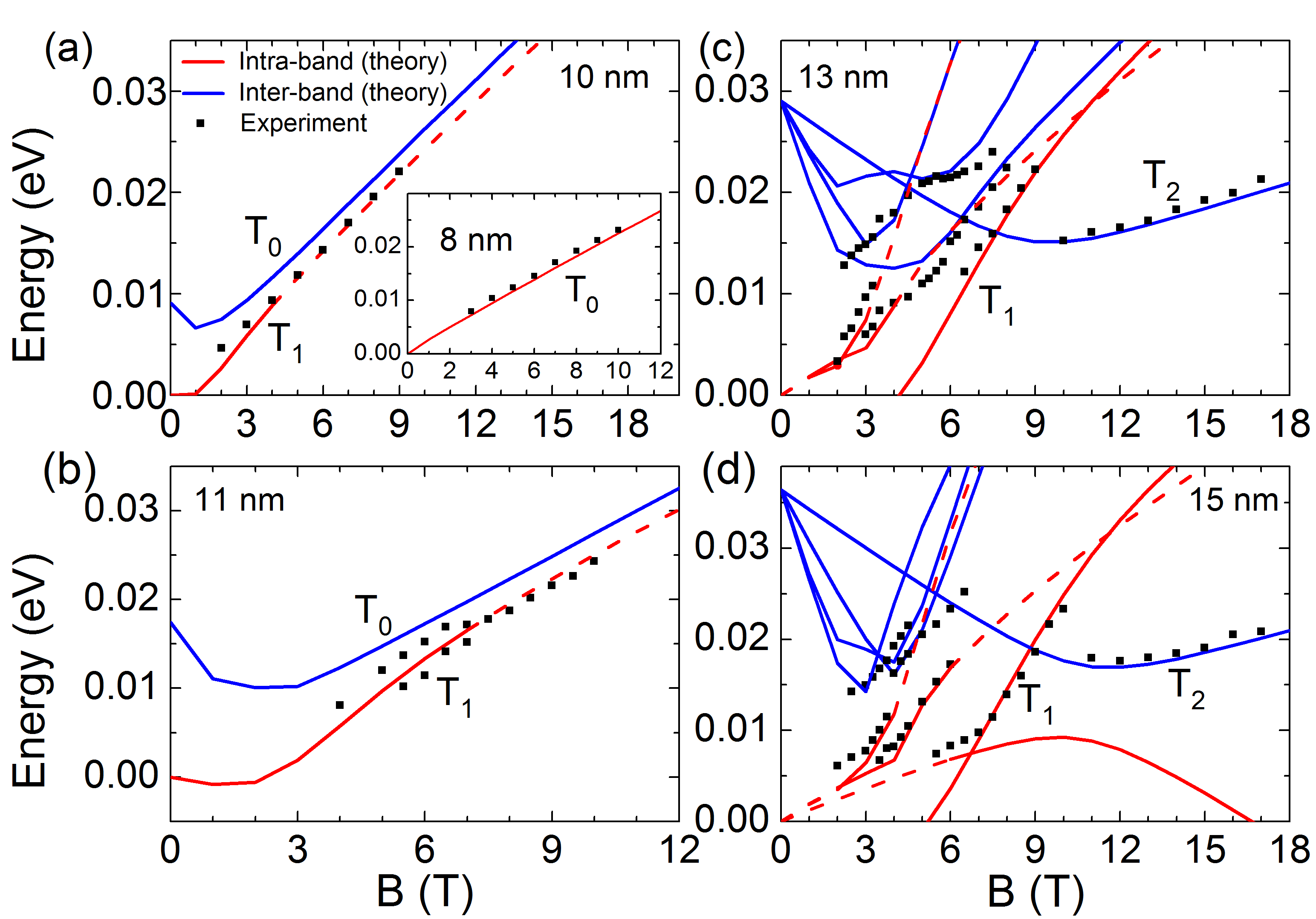}
\caption{(color online) Magnetic field dependence of major absorption peaks, both experimental and theoretical, for the $d=8$ nm (inset to (a)), 10 nm (a), 11 nm (b), 13 nm (c), and 15 nm (d) DQW samples. The blue (red) solid lines represent the manifold-resolved inter-band (intra-band) transitions. For simplicity, we only label the commonly observed transitions, $T_0$, $T_1$, and $T_2$, in our samples. The red dashed lines indicate the diminishing absorption peaks (Pauli) blocked when the corresponding level crosses above the Fermi energy.}
\end{figure}

It is important to note that our eight-band PB calculation is based on the band parameters reported in Ref. \cite{Vurgaftman2001} with only two adjustable variables. One is the carrier density (or Fermi energy), which is set to be $n=0.9 \times 10^{11}$ cm$^{-2}$ for all the samples. As our DQWs are close to the intrinsic limit, the Fermi energy is expected to be quickly pinned to the lowest electron level with increasing magnetic field. The fact that only one absorption peak ($T_0$) is observed in the normal state ($d=8$ nm) and slightly inverted ($d=10$ nm) samples suggests an upper bound for the electron density, $n \leq 0.9 \times 10^{11}$ cm$^{-2}$. The slight broadening of the CR linewidth in the $d=10$ nm sample at $B \leq 3$ T is indicative of the presence of the $T_1$ transition (Fig. 3(b)), which sets the lower bound for $n$. Therefore, one can conclude that $n=0.9 \times 10^{11}$ cm$^{-2}$ is a good approximation. The other variable used in our calculation is the effective mass of electrons in InAs QW, which is determined experimentally as $m_e^*=0.024m_0$ by fitting to the CR ($T_0$) peak in the normal state ($d=8$ nm) sample. Here, $m_0$ is the bare electron mass. In the inverted state, the contribution of $m_e^*$ to the band structure is coupled to that of the charge transfer effect discussed above. Therefore, $m_e^*$ can only be extracted accurately from the normal state data.

In addition to the $T_0$ transition, it is intriguing to investigate the $T_1$ (intra-band, $p:0\rightarrow1$) and $T_2$ (inter-band, $p:0\rightarrow-1$) transitions denoted in Figs. 3 and 4. We note that in the normal state, $T_1$ is indiscernible from the $T_0$ transition. It starts to depart from $T_0$ when the band is slightly inverted (Figs. 4(a,b)), therefore it is responsible for the asymmetric lineshape observed in the $d=11$ nm sample (Fig. S1). As the band is further inverted, $T_1$ becomes a well-developed peak (Figs. 4(c,d)), but only occurs when the magnetic field lifts the $p=1$ level (green) above the $p=0$ level (black), as shown in Figs. 3(c,d). Therefore, $T_1$ can be used as an effective band-inversion indicator, and practically one can perform a linear-in-$B$ fit to its energy at relatively high magnetic field, with the fitting intercept $y_0\approx0$ for the normal state, $y_0<0$ for the inverted state, and more negative value being more inverted. On the other hand, the $T_2$ transition only occurs in the inverted state and in the high-field region. It exhibits very weak $B$-dependence, distinct from all the other absorption peaks we have observed. We attribute this peak to a hole-like inter-band transition, $p:0\rightarrow-1$, due to the uplift of the heavy hole $p=-1$ level to the conduction band (\textit{i.e.}, band inversion). This observation is consistent with that reported in Ref. \cite{Nicholas_SST1}.

\begin{figure}[t]
\includegraphics[width=8.5cm]{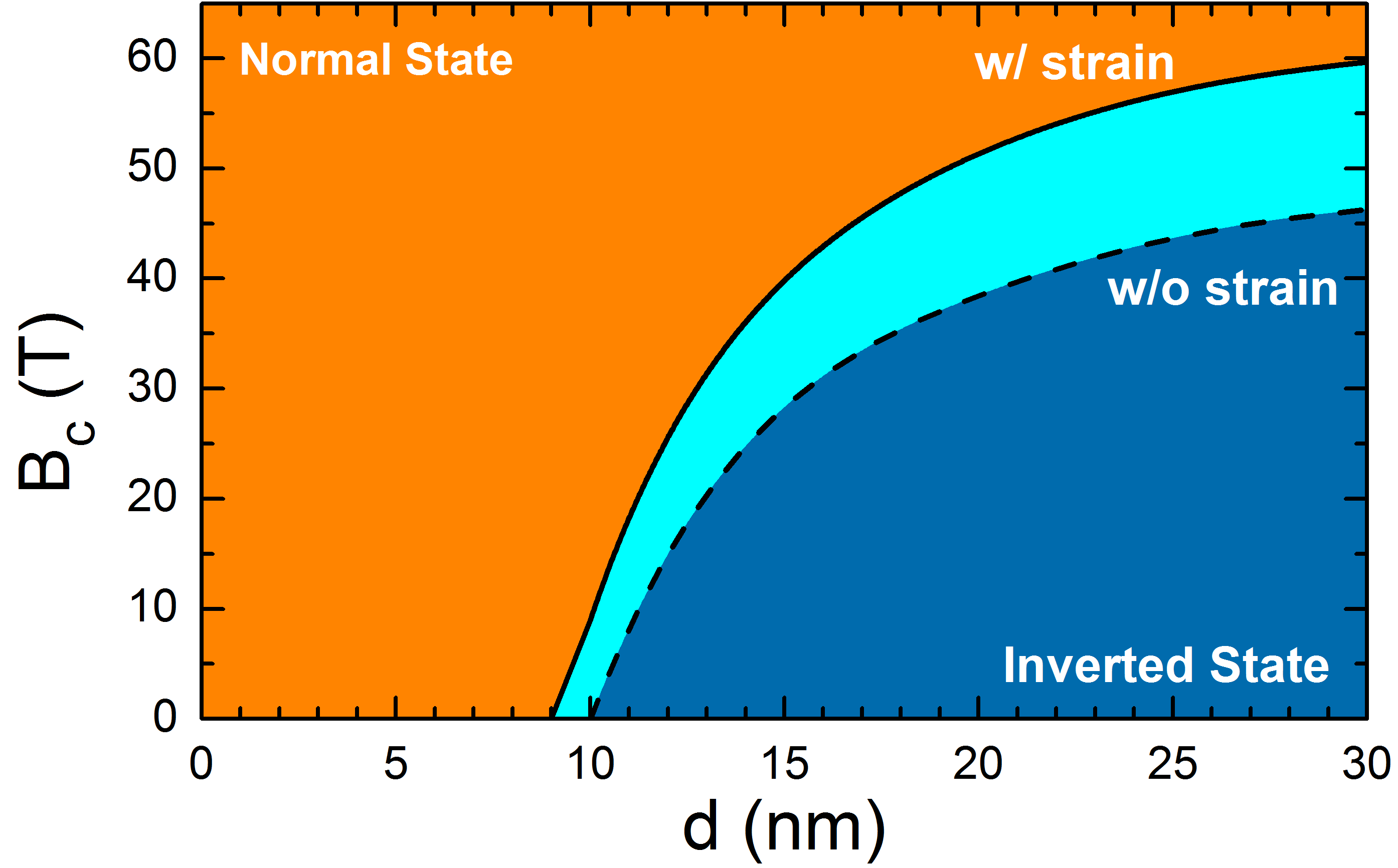}
\caption{(color online) Diagram describing the magnetic field driven semimetal to semiconductor transition from the inverted to the normal state, with and without strain. The width of GaSb QW is fixed to 5 nm. For demonstration purpose only, we omit the lengthy self-consistency calculation in this diagram.}
\end{figure}

Lastly, we return to the magnetic field driven semimetal to semiconductor transition from the inverted to the normal state. As mentioned above, the heavy hole level ($p=-1$), inverted to the conduction band, would eventually cross below the lowest electron level ($p=1$) with increasing magnetic field, driving the system back to the normal state. A critical field $B_c$ can be defined at this crossing point (Figs. 3(b,c)), and the corresponding transition diagram is plotted in Fig. 5. The importance of the strain effect can also be seen in Fig. 5, as it significantly shifts the boundary between the inverted state and the magnetic field driven normal state. Practically, the strain effect can be engineered by choosing the appropriate substrate (typically, GaSb for pseudomorphic growth, GaAs for metamorphic growth) and the epi-structure \cite{Du_new}.  

In conclusion, we have studied the LL structure of a series of InAs/GaSb DQWs from the normal to the inverted state using magneto-IR spectroscopy. We find that close to the intrinsic limit, the band inversion significantly modifies the magneto-absorption of the system, giving rise to multiple absorption peaks with distinct non-linear $B$-dependence. All the major absorption peaks observed in our experiment can successfully be explained using an eight-band PB model, with semi-quantitative agreement surpassing the previous two-band \cite{Nicholas_PRB1} and six-band \cite{Nicholas_SST1} models. The application of our model to InAs/GaSb multilayers and superlattices is currently in progress.

We thank Danhong Huang for helpful discussion. This work was primarily supported by the DOE (Grant No. DE-FG02-07ER46451). S.T., G.D.S., and C.J.S. acknowledge support from the NSF (Grant No. DMR-1311849) and the AFOSR (Grant No. FA9550-14-1-0376). Q.Z. and J.K. acknowledge support from the NSF (Grant No. DMR-1310138). The work at SNL is supported by the DOE Office of Basic Energy Sciences, Division of Materials Science and Engineering, and by Sandia LDRD. Sandia National Laboratories is a multi-program laboratory managed and operated by Sandia Corporation, a wholly owned subsidiary of Lockheed Martin Corporation, for the United States Department of Energy's National Nuclear Security Administration under contract DE-AC04-94AL85000. The magneto-IR measurements were performed at the NHMFL, which is supported by the NSF Cooperative Agreement No. DMR-1157490 and the State of Florida.


\begin{thebibliography}{99}
\bibitem{note1}We stress that the semi-quantitative agreement is reached by using one set of parameters for five InAs/GaSb DQW samples across the normal and inverted states.
\end{thebibliography}
\end{document}


\title{Supplementary Material: Probing the semiconductor to semimetal transition in InAs/GaSb double quantum wells by magneto-infrared spectroscopy}

\author{Y. Jiang}
\affiliation{School of Physics, Georgia Institute of Technology, Atlanta, Georgia 30332, USA}
\author{S. Thapa}
\affiliation{Department of Physics, University of Florida, Gainesville, Florida 32611, USA}
\author{G. D. Sanders}
\affiliation{Department of Physics, University of Florida, Gainesville, Florida 32611, USA}
\author{C. J. Stanton}
\affiliation{Department of Physics, University of Florida, Gainesville, Florida 32611, USA}
\author{Q. Zhang}
\affiliation{Department of Electrical and Computer Engineering, Department of Physics and Astronomy, and Department of Materials Science and NanoEngineering, Rice University, Houston, Texas 77005, USA}
\author{J. Kono}
\affiliation{Department of Electrical and Computer Engineering, Department of Physics and Astronomy, and Department of Materials Science and NanoEngineering, Rice University, Houston, Texas 77005, USA}
\author{W. K. Lou}
\affiliation{SKLSM, Institute of Semiconductors, Chinese Academy of Sciences, Beijing 100083, China}
\affiliation{Synergetic Innovation Center of Quantum Information and Quantum Physics, University of Science and Technology of China, Hefei, Anhui 230026, China}
\author{K. Chang}
\affiliation{SKLSM, Institute of Semiconductors, Chinese Academy of Sciences, Beijing 100083, China}
\affiliation{Synergetic Innovation Center of Quantum Information and Quantum Physics, University of Science and Technology of China, Hefei, Anhui 230026, China}
\author{S. D. Hawkins}
\affiliation{Sandia National Laboratories, Albuquerque, New Mexico 87185, USA}
\author{J. F. Klem}
\affiliation{Sandia National Laboratories, Albuquerque, New Mexico 87185, USA}
\author{W. Pan}
\affiliation{Sandia National Laboratories, Albuquerque, New Mexico 87185, USA}
\author{D. Smirnov}
\affiliation{National High Magnetic Field Laboratory, Tallahassee, Florida 32310, USA}
\author{Z. Jiang}
\email{zhigang.jiang@physics.gatech.edu}
\affiliation{School of Physics, Georgia Institute of Technology, Atlanta, Georgia 30332, USA}

\date{\today}

\maketitle

\section{I. Additional Experimental Results}
Figure S1(a) shows normalized magneto-absorption spectra, $-T(B)/T(B = 0)$, for the $d=11$ nm double quantum well (DQW) sample at selected magnetic fields. Here, the spectra exhibit a much broader and asymmetric lineshape, particularly below 8 T, compared with that of the $d=10$ nm sample shown in Fig. 2(a) of the main text. Our calculation in Fig. S1(b) suggests that there exist two absorption peaks at low magnetic field. The lower energy peak vanishes around 9 T due to Pauli blocking when the corresponding energy level crosses above the Fermi energy.

\begin{figure}[b]
\centering
\includegraphics[width=13cm]{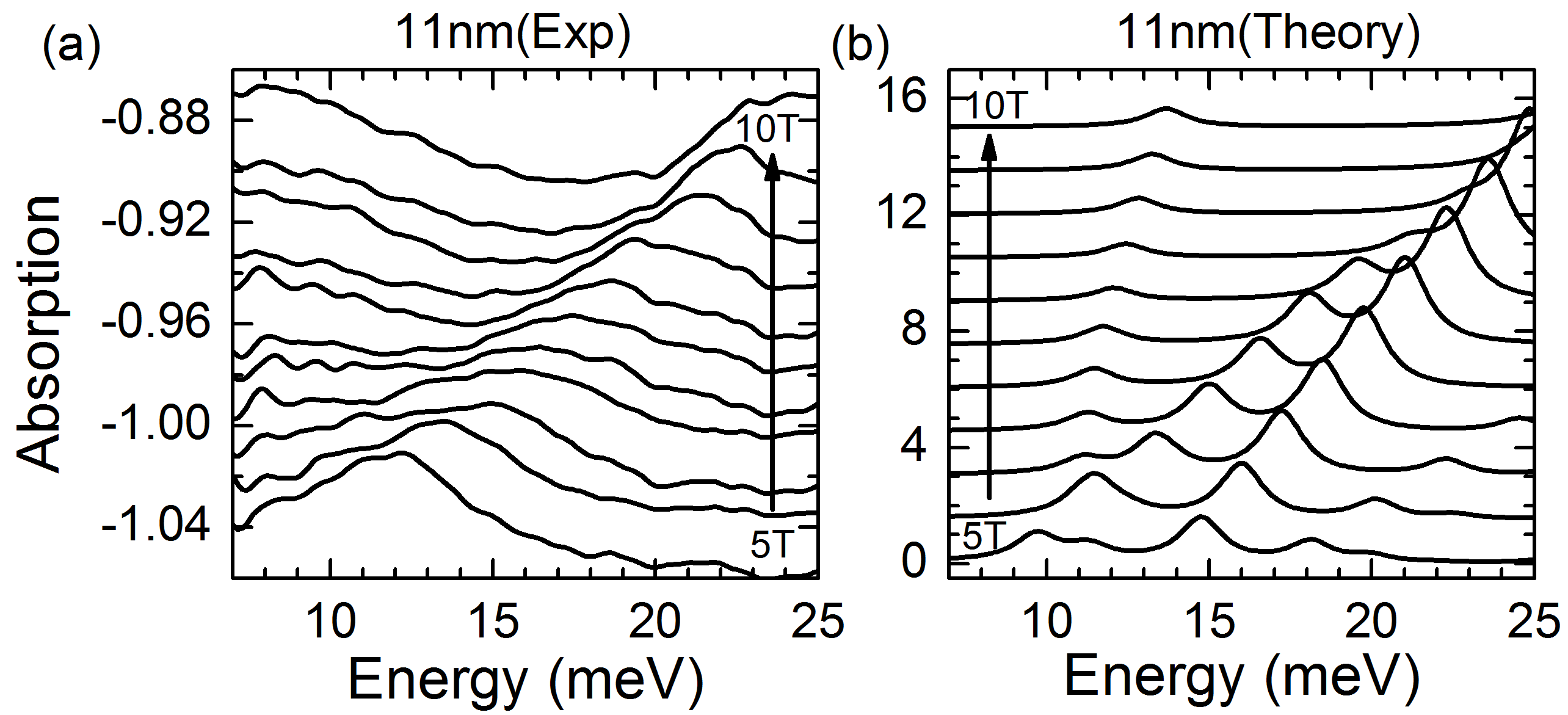}
\caption{(a) Normalized magneto-absorption spectra, $-T(B)/T(B = 0)$, for the $d=11$ nm DQW sample at selected magnetic fields. (b) Calculated magneto-absorption spectra using eight-band Pidgeon-Brown (PB) model assuming a FWHM of 0.8 meV. In both panels, the spectra are offset vertically for clarity.}
\end{figure}

\begin{figure}[t]
\includegraphics[width=12.25cm] {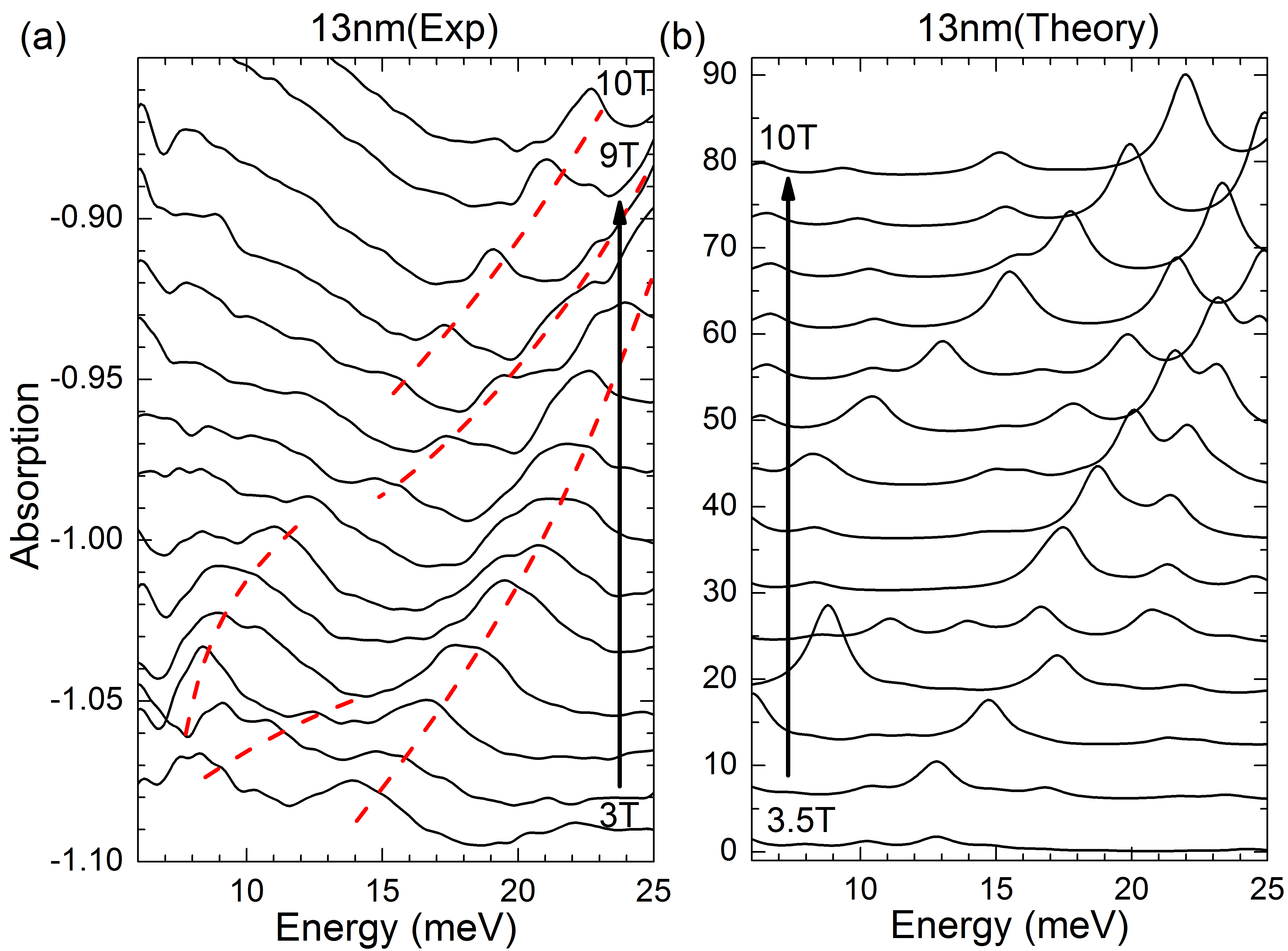}
\caption{(color online) (a) Normalized magneto-absorption spectra, $-T(B)/T(B = 0)$, for the $d=13$ nm DQW sample at selected magnetic fields. The dashed lines indicate the major absorption peaks observed in the experiment. (b) Calculated magneto-absorption spectra using eight-band PB model assuming a FWHM of 0.8 meV. In both panels, the spectra are offset vertically for clarity.}
\end{figure}
 
In Fig. S2, we plot normalized magneto-absorption spectra of the $d=13$ nm DQW sample and compare with the theoretical calculation. In the low-field region ($B\le5$ T), one can clearly identify three absorption peaks (indicated by dashed lines). The two lower energy peaks are cyclotron resonance (CR)-like, while the higher energy one may be attributed to an inter-band transition. With increasing magnetic field, the two CR-like peaks vanish, again due to Pauli blocking, and another two peaks appear in the intermediate-field region (up to 10 T) with one CR-like and the other related to inter-band transition, as shown in Fig. 4(c) of the main text. 

\section{II. Additional Energy Level Diagram}
Figure S3 shows the calculated energy levels and Fermi energy $E_F$ as a function of magnetic field for the $d = 11$ nm DQW sample. Like in Fig. 3 of the main text, the PB manifold is color coded based on the index $p$, and for simplicity, we only show the lowest levels, \textit{i.e.}, when $p=-1$, 0, 1, and 2.
 
\begin{figure}[t]
\includegraphics[width=8.5cm] {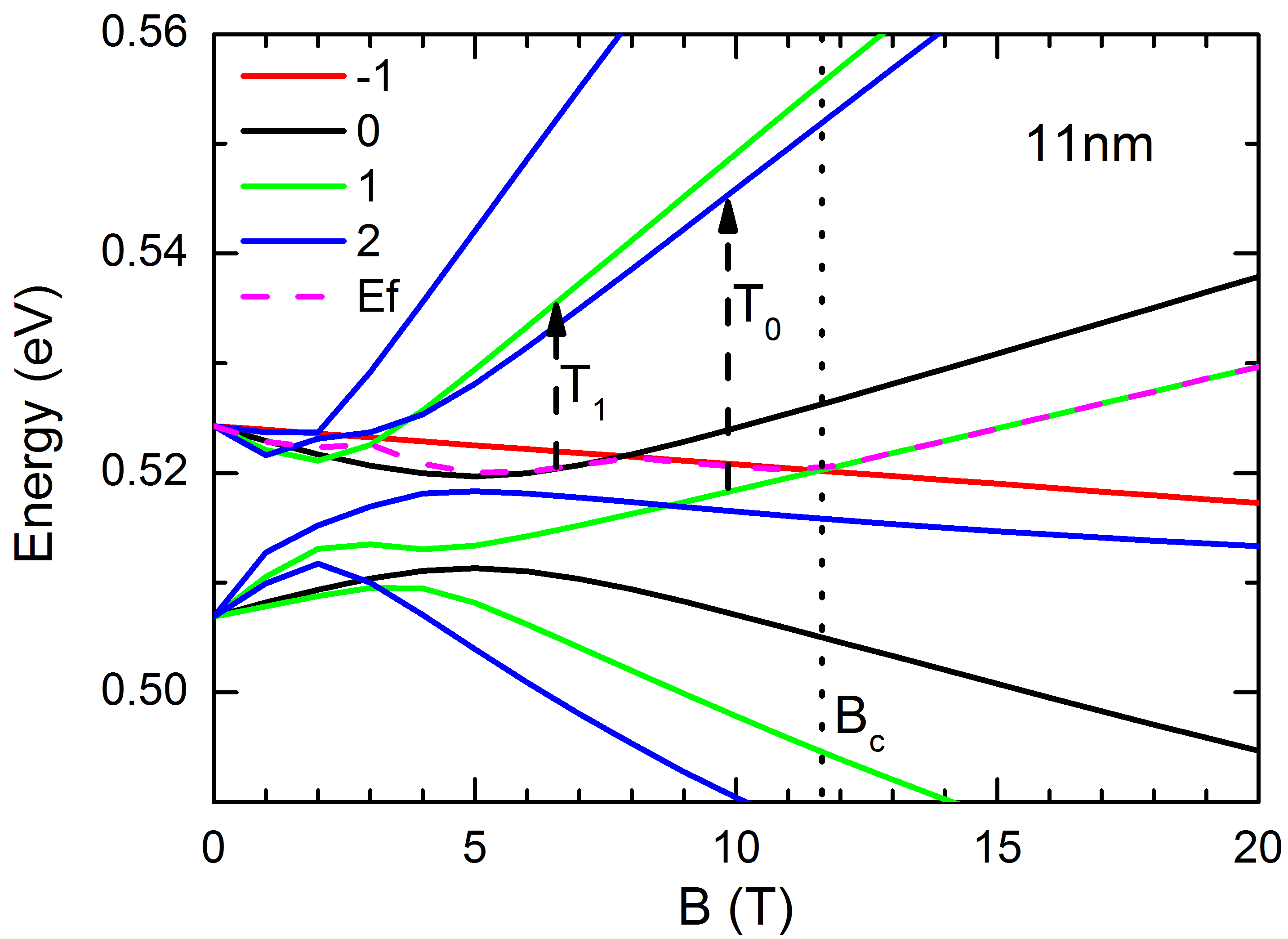}
\caption{(color online) Calculated energy levels as a function of magnetic field for the $d = 11$ nm DQW sample. The PB manifold is color coded based on the index $p$, and the dashed line shows the evolution of $E_F$ as a function of magnetic field. The dashed arrows indicate the two major transitions, $T_0$ and $T_1$, commonly observed in our samples. The dotted line marks the onset ($B_c$) of the magnetic field driven transition from the inverted to the normal state.}
\end{figure}

\section{III. Eight-band $\mathbf{K\cdot P}$ Calculation}
In this Section, we briefly describe the $\mathbf{k\cdot p}$ calculation used to obtain the electronic band structure of InAs/GaSb DQWs at zero magnetic field. First, we start with the Kane Hamiltonian for bulk zinc-blende semiconductors and the corresponding basis functions, as denoted in Ref. \cite{Kai_PRB1}. We assume quantum confinement effect along the $z$-direction, and replace the $k_z$ wave-vector component in the Kane Hamiltonian by momentum operator $-i\partial/\partial_z$. The in-plane wave vector $(k_x, k_y)$, on the other hand, remains continuous due to translational invariance. Second, we follow the envelope function theory of Ref. \cite{Burt1992} and expand the wavefunction $\Psi(\bold{r})$ in terms of the envelope function $F_n$ and the Bloch function $U_n$ of the $n^{th}$ spinor
\begin{align*}
\Psi(\bold{r})=\sum_{n}{F_n(\bold{r})U_n(\bold{r})},
\end{align*}
where $F_n(\bold{r})=\exp(ik_x x+ik_y y) f_n (z)$ and $U_n$ is assumed to be material independent. We then solve the Schr\"{o}dinger equation with this set of envelope functions by expanding $f_n (z)$ in plane wave basis. For the $s^{th}$ subband,
\begin{align*}
f^s_n(z)=\frac{1}{\sqrt{L}}\sum_{m=-N}^{N}c_{nm}^s\exp(ik_mz),
\end{align*}
where $L$ is the total width of the AlSb/InAs/GaSb/AlSb quantum well (QW) structure and $k_m=2m\pi�/L$. In our calculation, we use $N=30$ to obtain convergent results while avoiding spurious solutions. For simplicity, we assume all the band parameters are discontinuous at each interface and can be described by a piecewise function along the growth direction. We also take the zero temperature limit and the axial approximation in our calculation.

\begin{table}[t]
\begin{center}
\begin{tabular}{| p{3cm} |l | c  |c   | c | }
\hline 
\hline 
 & InAs & GaSb & AlSb \\
\hline
$m^*_e$($m_0$)&0.024&0.039&0.14\\
$E_g$(eV) & 0.417 & 0.812 & 2.386 \\
$E_v$(eV) & -0.417 & 0.143 & -0.237  \\
$E_p$(eV) & 21.5 & 27 & 18.7 \\
$\Delta$ (eV) & 0.39 & 0.76 & 0.676  \\
$A_c$(eV $nm^2$) & -0.0601 & -0.0853 & -0.0045  \\
$\gamma_1$ & 2.81& 2.32 & 2.57  \\
$\gamma_2$ & -0.093 & -0.842 & -0.116 \\
$\gamma_3$ & 0.607 & 0.458 & 0.664\\
$\kappa$ & -1.060 & -1.542 & -0.936\\
$a$(\AA) & 6.050 & 6.082 & 6.128\\
$C_{12}(GPa)$ & 452.6 & 402.6 & 434.1\\
$C_{11}(GPa)$ & 832.9 & 884.2 & 876.9\\
$a_c$(eV) & -5.08 & -7.5 & -4.5\\
$a_v$(eV) & -1 & -0.8 & -1.4\\
$b$(eV) & -1.8 & -2.0 & -1.35\\
$d$(eV) & -3.6& -4.7 & -4.3\\
\hline
\end{tabular}
\end{center}
\caption{Band parameters used in our eight-band $\mathbf{k\cdot p}$ calculation. Unless stated explicitly, the parameters are taken from Ref. \cite{Vurgaftman2001}.} 
\end{table}

Table S1 summarizes the band parameters used in our calculation. Except for the $m^*_e$ and $A_c$ values in InAs (the latter is a function of $m^*_e$), the band parameters are taken from Ref. \cite{Vurgaftman2001}. As discussed in the main text, the effective mass of electrons in InAs QW, $m^*_e= 0.024m_0$, is determined experimentally from our data.

\justify
\textit{\textbf{Determining the Fermi Energy.}} For an electron-hole hybridized system (inverted state), the $E_F$ calculation deserves special attention. In our case, we do not distinguish the electron and the hole composition in a specific state/level. Instead, we simply assign the lowest six bands within the eight-band model valence bands and the highest two bands conduction bands. After the plane wave expansion, we label the lowest $6\times(2N+1)$ bands valence bands and the highest $2\times(2N+1)$ bands conduction bands. With this labeling in addition to electron and hole Fermi-Dirac statistics, we are then able to determine $E_F$ for a given carrier density. 

\justify
\textit{\textbf{Self-consistent potential profile calculation for the inverted state.}} In a broken-gap heterostructure, the charge transfer across the interface may significantly modify the internal electrostatic potential and hence the band structure. In principle, the potential profile $V(z)$ can be calculated by solving the Poisson equation with a Neumann boundary condition using the electron/hole distribution deduced from the envelope functions \cite{Kai_PRB1}. Then, one can iterate the calculated $V(z)$ in the Schr\"{o}dinger equation until convergence is reached. However, in practice the $V(z)$ calculation is coupled with the $E_F$ calculation because the charge neutrality condition must remain satisfied during the charge transfer. Here, we decouple these two calculations using the following method. First, we determine the $E_F$ as described above. Then, we calculate the electron(hole) component that is inverted into the valence(conduction) band. Next, we apply the charge neutrality condition and scale the hole component with respect to the electron component. Finally, we calculate $V(z)$. We repeat this calculation for each iteration. In addition, following Ref. \cite{Chao_PRB2} we further assume that the self-consistent potential profile calculated at zero magnetic field is $B$-field independent. In this way, we can extend the eight-band model to include the magnetic field effect.

\section{IV. Pidgeon-Brown index}
Using Peierls substitution with the choice of magnetic vector potential $\bold{A}=(0,Bx,0)$, one can choose to express the envelope functions as \cite{Buhmann_PRB1}
\[
F_{p,\nu}=\exp{(ik_yy)}
  \begin{bmatrix}
    f_{p,1,\nu}(z) \phi_{p-1} \\f_{p,2,\nu}(z) \phi_{p}\\ f_{p,3,\nu}(z) \phi_{p-2} \\f_{p,4,\nu}(z) \phi_{p-1} \\ f_{p,5,\nu}(z) \phi_{p}\\ f_{p,6,\nu}(z) \phi_{p+1}\\ f_{p,7,\nu}(z) \phi_{p-1}\\f_{p,8,\nu}(z) \phi_{p}
  \end{bmatrix},
\]
where translational symmetry in the $x$-direction is broken while $k_y$ remains as a good quantum number. $\phi_p$ is the $p^{th}$ harmonic oscillator eigenfunction, $p$ is the PB index, and the integer $\nu$ labels the energy levels within the same PB manifold in an ascending order. We note that the harmonic oscillator eigenfunction $\phi_p$ is only well defined for $p\ge0$ (one can assume $\phi_p=0$ when $p<0$), so the PB index runs from $-1$ to infinity, \textit{i.e.}, $p=-1, 0, 1, 2 ...$, for non-zero wavefunctions.

\section{V. Manifold-Resolved Magneto-Absorption Spectrum}
Once the wavefunctions are obtained, one can calculate the magneto-absorption spectrum from the imaginary part of the dielectric function $\epsilon_2$ deduced using Fermi's golden rule \cite{Stanton_PRB1}. The magneto-absorption coefficient $\alpha$ at the photon energy $\hbar \omega$ is then given by
\begin{align*}
\alpha(\hbar \omega)&=\frac{\omega}{cn_r}\epsilon_2(\hbar\omega)\\
&\propto\frac{e^3 B}{\hbar \omega (\hbar c)^2 n_r} \sum_{m,n,m',n'} |\hat{\bold{e}} \cdot \vec{\bold{P}}_{m,n}^{m',n'}|^2\times (f_{m,n}-f_{m',n'}) \delta(\Delta E_{m,n}^{m',n'}-\hbar \omega),
\end{align*}
where $\hbar$ is reduced Planck's constant, $c$ is the speed of light, $n_r$ is the index of refraction, $e$ is the electron charge, $(m',n')$ and $(m,n)$ denote the final and initial states, $\hat{\bold{e}}$ is the unit polarization vector, $\hat{\bold{e}} \cdot \vec{\bold{P}}_{m,n}^{m',n'}$ are the optical matrix elements (detailed in Ref. \cite{Stanton_PRB1}), $f_{m,n}$ is the Fermi distribution, and $\Delta E_{m,n}^{m',n'}$ is the optical transition energy. To compare with experimental data, the Dirac Delta function is replaced by a Lorentzian function with a full width at half maximum (FWHM) of 0.8 meV extracted from the measured spectra.

By inspecting the optical matrix, one can find that only transitions with PB manifolds that differe by $\Delta p = \pm 1$ are allowed. Specifically, the $p \rightarrow p+1 $ transition is allowed for $\sigma^+$ polarized light, while $p \rightarrow p-1 $ is for $\sigma^-$ polarized light. Therefore, $\Delta p = \pm 1$ are associated with the electron- and the hole-like transitions, respectively. The total absorption can then be decomposed into a series of manifold-resolved absorption peaks, originating from a specific manifold $p$ with either $\Delta p=+1$ or $\Delta p=-1$. Some manifold-resolved magneto-absorption spectra are plotted in Fig. S4 for the $d=15$ nm DQW sample, following the normalization method used in Fig. 2 of the main text. The calculations are compared with the major absorption peaks observed in the experiment (red): two major peaks in panel (a), three in panel (b), and one each for panels (c) and (d). Taking Fig. S4(b) as an example, the high-energy peak can be attributed to a $p: 0 \rightarrow 1$ electron-like transition, whereas the two low-energy peaks are from the hole-like transitions originating from $p=2$, 1 and 0. The lowest peak may consist of contributions from two different transitions, $p: 1 \rightarrow 0$ and $p: 0 \rightarrow -1$. Similar analyses are performed on all the magneto-absorption spectra obtained from the five samples and the results are summarized in Fig. 4 of the main text.

\begin{figure}[t]
\includegraphics[width=14cm] {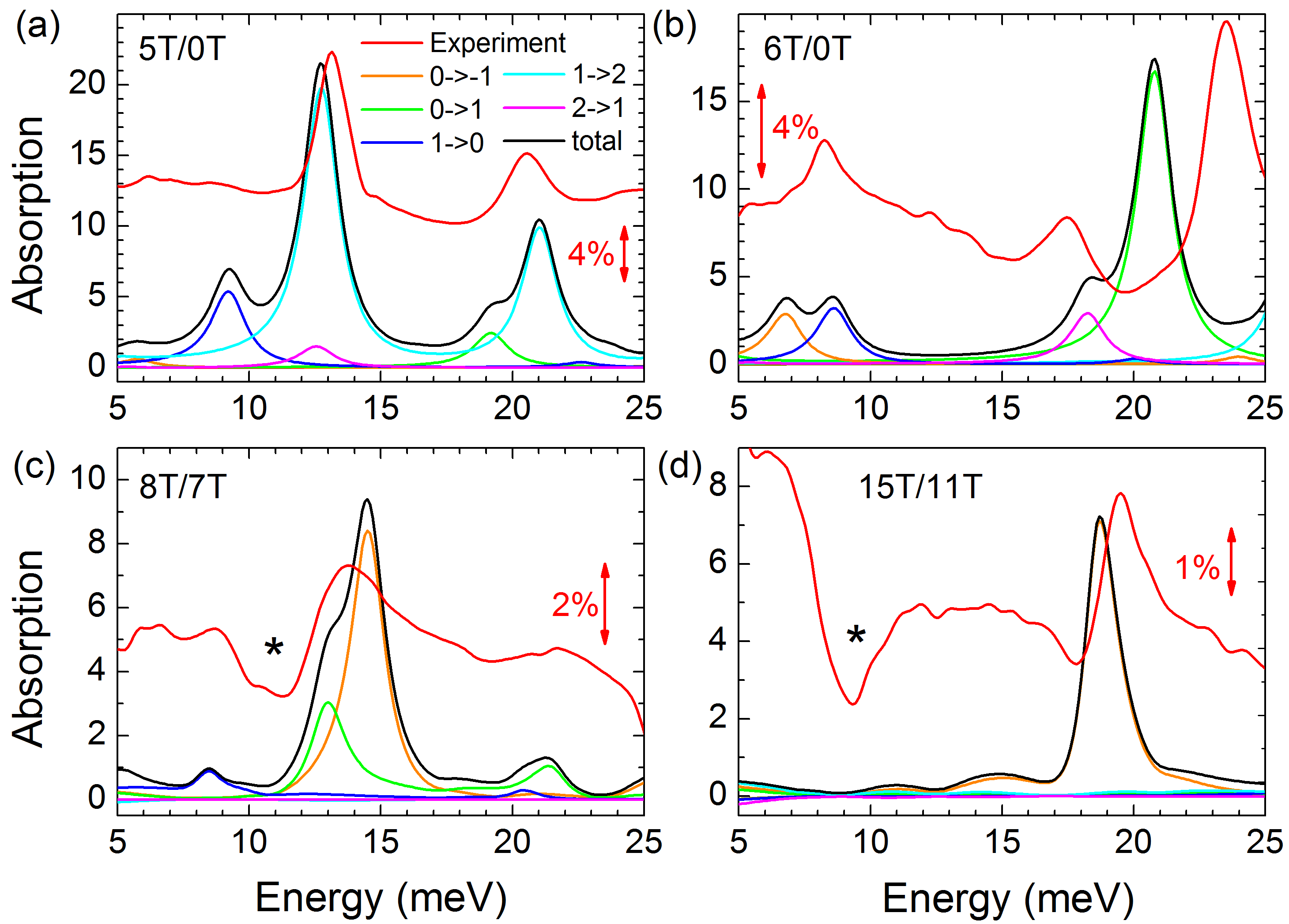}
\caption{(color online) Calculated manifold-resolved magneto-absorption spectra for the $d=15$ nm DQW sample at selected magnetic fields. The manifold-resolved absorptions are color coded, while the total absorption is in black. The star symbols point to $B$-independent spectral features (dips) resulting from the normalization process. The calculations are compared with the major absorption peaks observed in the experiment (red), following the normalization method used in Fig. 2 of the main text. Specifically, panels (a,b) are normalized to zero field, while panel (c) and (d) are normalized to 7 T and 11 T, respectively.}
\end{figure}